\documentclass[twocolumn,showpacs,english,superscriptaddress,aps,prl,preprintnumbers,amsmath,amssymb,floatfix]{revtex4-1}

\usepackage[T1]{fontenc}
\usepackage[latin9]{inputenc}
\usepackage{dcolumn}
\usepackage{bm}
\usepackage{graphicx}
\usepackage{color}
\usepackage{esint}
\usepackage{babel}
\usepackage{amsfonts}
\usepackage{slashed}
\usepackage{enumerate}
\usepackage{color}
\usepackage[bookmarks=true,colorlinks,linkcolor=blue,urlcolor=blue,citecolor=blue]{hyperref}

\newcommand{\be}{\begin{equation}}
\newcommand{\ee}{\end{equation}}
\newcommand{\bea}{\begin{eqnarray}}
\newcommand{\eea}{\end{eqnarray}}

\newcommand{\mc}{\mathcal}

\begin{document}

\title{Quantum  spin circulator in  Y junctions of Heisenberg chains}

\author{Francesco Buccheri}
\affiliation{Institut f\"ur Theoretische Physik,
Heinrich-Heine-Universit\"at, D-40225  D\"usseldorf, Germany}

\author{Reinhold Egger}
\affiliation{Institut f\"ur Theoretische Physik,
Heinrich-Heine-Universit\"at, D-40225  D\"usseldorf, Germany}

\author{Rodrigo G.~Pereira}
\affiliation{International Institute of Physics, Universidade Federal do Rio Grande do Norte, Campus Universitario, Lagoa Nova, Natal-RN 59078-970, Brazil}

\author{Fl\'avia B. Ramos}
\affiliation{International Institute of Physics, Universidade Federal do Rio Grande do Norte, Campus Universitario, Lagoa Nova, Natal-RN 59078-970, Brazil}

\date{\today}

\begin{abstract}
We show that a quantum spin circulator, a nonreciprocal device that routes spin currents without any charge transport, can be achieved in Y junctions of identical spin-$1/2$ Heisenberg chains coupled  by a chiral three-spin interaction. Using bosonization, boundary conformal field theory, and  density-matrix renormalization group simulations, we find that a chiral fixed point with maximally asymmetric spin conductance arises at a critical point separating  a regime of disconnected chains  from  a spin-only version of the three-channel Kondo effect. We argue that networks of spin-chain Y junctions
provide a controllable approach to construct long-sought chiral spin liquid phases. 

\end{abstract}

\maketitle
\emph{Introduction.---}The spin-$1/2$ Heisenberg chain represents an
analytically accessible model of basic importance in condensed matter theory
\cite{Gogolin1998}. By now, many experimental and theoretical works have contributed to 
a rather complete understanding of this model, including the effects of boundaries and junctions of two chains \cite{Eggert1992}. However, 
little attention has been devoted to quantum junctions formed by more than two
Heisenberg chains. In fact, recent theoretical developments provide hints 
that interesting physics should be expected in that direction:  First, 
multichannel   Kondo fixed points have been predicted for 
junctions of  anisotropic spin chains \cite{Tsvelik2013,Crampe2013,Tsvelik2013b,Buccheri2015}.
Second, electronic charge transport through junctions of three quantum wires is governed by
a variety of nontrivial fixed points which cannot be realized in two-terminal setups
\cite{Nayak1999,Chen2002,Chamon2003,Barnabe2005,Oshikawa2006,Hou2008,Giuliano2009,Agarwal2009,Bellazzini2009,
Rahmani2012}.
As spin currents  in antiferromagnets can be induced  by spin pumping
\cite{Cheng2014} or by the longitudinal spin-Seebeck effect \cite{Hirobe2016},
it is both an experimentally relevant and fundamental question to determine
nontrivial fixed points governing spin transport in junctions of
multiple spin chains. In particular, we are interested in the possibility of
realizing a circulator for spin currents.  While circulators have been discussed for photons
\cite{Scheucher2016,Lodahl2016,Chapman2017} and for quantum Hall edge states
\cite{Viola2014,Mahoney2017}, we are not aware of existing proposals for spin
circulators. Once realized, a spin circulator has immediate applications in the field of spintronics
\cite{Wolf2001}, which has recently turned to the study of charge-insulating antiferromagnetic materials \cite{Wadley2016,Jungwirth,Baltz2016}.

In this paper, we  study Y junctions of  spin-$1/2$ Heisenberg
chains coupled at their ends by spin-rotation [SU(2)] invariant interactions.
We assume identical chains such that the junction is $\mathbb{Z}_3$-symmetric under 
a cyclic exchange. These conditions are respected by a chiral three-spin coupling $J_\chi$
[see Eq.~\eqref{couplH} below], which breaks time reversal (${\cal T}$) symmetry and  
can be tuned from weak to strong coupling, e.g., by changing an Aharonov-Bohm flux
 \cite{Wen1989,Sen1995,Claassen2017}.  Apart from condensed matter systems, 
such Y junctions can also be studied in ultracold atom platforms \cite{Esslinger2010}, 
where Heisenberg   chains  \cite{Murmann2015,Boll2016,Endres2016} and multi-spin exchange 
processes \cite{Dai2017} have recently been realized.
We use three complementary theoretical approaches, namely   
bosonization \cite{Gogolin1998},  boundary conformal field theory (BCFT) \cite{Cardy1986,Cardy1989,Affleck1991,AffleckLudwig2,Affleck1993}, 
and density matrix renormalization group (DMRG) simulations \cite{White1992,Schollwock2005}.     

\begin{figure}[b]
\centering
\includegraphics[width=0.98\columnwidth]{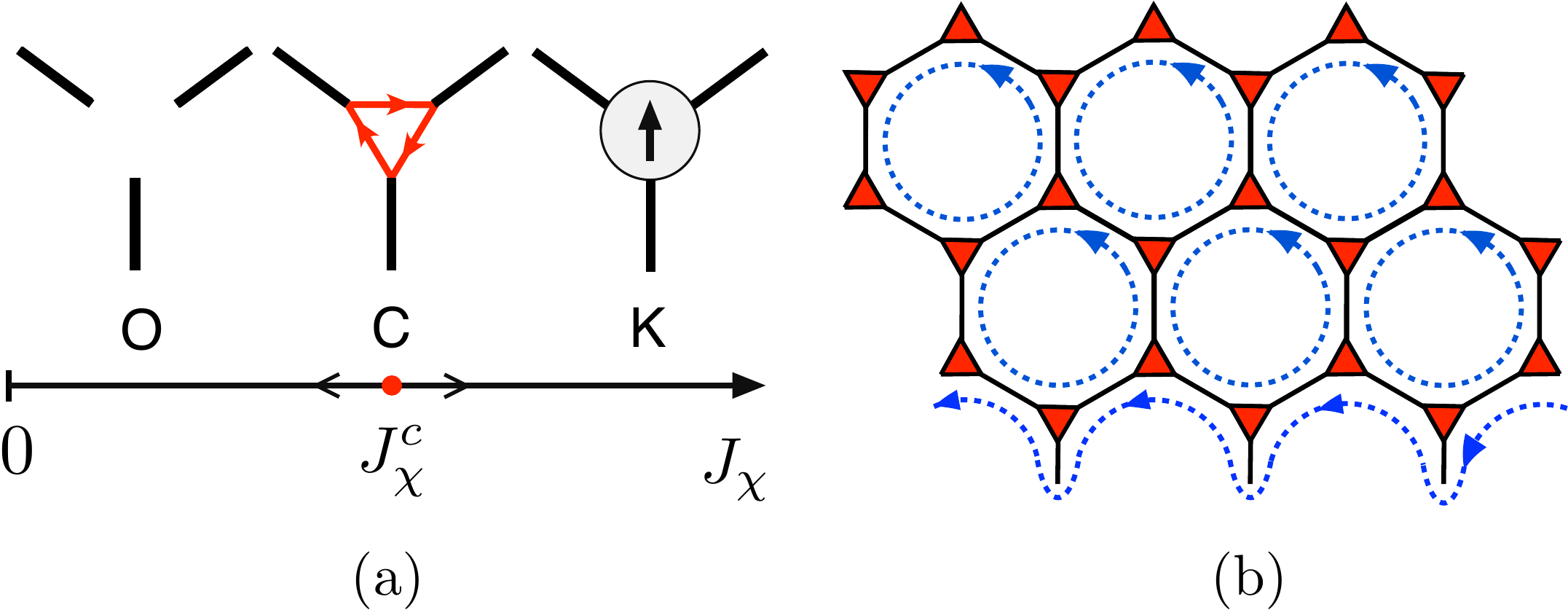}
\caption{Y junction and network. (a) Schematic illustration of the phase diagram.   For $J_\chi<J_\chi^c$, the
system flows to open boundary conditions (O fixed point), while 
for $J_\chi>J_\chi^c$, the three-channel Kondo (K) point  is approached.
The two stable fixed points are separated by an unstable chiral (C) fixed point
at $J_\chi=J_\chi^c$. (b) A network of Y junctions with uniform $J_\chi$ tuned to the C point  realizes a chiral spin liquid. }
\label{fig1}
\end{figure}

Before entering a detailed discussion, we briefly describe our main 
conclusions, see Fig.~\ref{fig1}(a):
(i) We find two stable fixed points with  emergent ${\cal T}$ symmetry. 
For small $|J_\chi|$, the renormalization group (RG) flow 
is towards the fixed point of open boundary conditions (O) representing disconnected chains.
For large $|J_\chi|$, however,  the system flows towards a spin-chain version of the three-channel Kondo fixed point 
\cite{Affleck1991}, referred to as K point in what follows.  So far only the 
two-channel Kondo effect with spin chains has been studied 
\cite{Eggert1992,Affleck1993,Alkurtass2016}.  
(ii) Both stable points are separated by an unstable chiral 
fixed point at intermediate coupling $|J_\chi|=J_\chi^c$, where
the circulation sense 
is determined by the sign of $J_\chi$.  DMRG simulations give $J^c_\chi/J=3.11(1)$, where $J>0$ is the bulk exchange coupling. 
(iii) Although the chiral point is unstable, it determines the physics over a wide  regime of intermediate values of $J_\chi$. 
It then realizes an ideal spin circulator, 
where incoming spin currents are scattered in a chiral 
(left- or right-handed) manner around the Y junction.  
 (iv) These findings provide a key step towards realizing a chiral spin liquid (CSL), an exotic  phase of frustrated quantum magnets   \cite{Kalmeyer1987,Wen1989,Bauer2013arXiv,Bauer2014,He2014,Gong2014,
 Gorohovsky2015,Bieri2015,Kumar2016}. 
Our spin circulator provides a building block for network constructions of CSLs,
cf.~Fig.~\ref{fig1}(b), where the chirality of each Y junction can be individually addressed.  

\emph{Model.---}We employ the Hamiltonian $H=H_0+H_c$, where 
$H_0=\sum_{j,\alpha} (J{\bf S}_{j,\alpha}\cdot {\bf S}_{j+1,\alpha}+J_2{\bf S}_{j,\alpha}\cdot {\bf S}_{j+2,\alpha})$
describes three ($\alpha=1,2,3$) identical semi-infinite Heisenberg
chains (lattice sites $j=1,2,\ldots$). In numerical studies, it is convenient to   tune the next-nearest-neighbor coupling $J_2=0.2412J$   to suppress  
logarithmic corrections present for $J_2=0$ \cite{Eggert1992}.  
The part $H_c$ captures couplings between the 
boundary spin-$1/2$ operators ${\bf S}_\alpha\equiv {\bf S}_{j=1,\alpha}$.  
We require $H_c$ to preserve spin-SU(2) invariance and $\mathbb{Z}_3$ symmetry under 
cyclic chain exchange, $\alpha\to \alpha+1$ with ${\bf S}_{\alpha=4}= {\bf S}_1$.  
These conditions allow for a ${\cal T}$-breaking three-spin coupling $J_\chi$,  
\begin{equation} \label{couplH}
H_c= J_\chi \hat C  , \qquad
\hat C= {\bf S}_{1}\cdot ({\bf S}_2\times {\bf S}_3),
\end{equation}
where $\hat C$ is the scalar spin chirality of the boundary spins \cite{Wen1989}.  We note that $J_\chi$ breaks reflection (${\cal P}$) symmetry,
 defined as exchange of chains $1$ and $2$, but  
$H$ is invariant under the composite ${\cal P T}$ symmetry. 
The $J_\chi$ interaction could be realized as an effective Floquet spin model for Mott insulators pumped by circularly polarized light \cite{Claassen2017}. In principle, the ratio $J_\chi/J$ can be made arbitrarily large by varying bulk and boundary parameters independently.  
 The above symmetries also allow for a $\mc T$-invariant boundary exchange coupling term, $J' \sum_\alpha {\bf S}_{\alpha} \cdot {\bf S}_{\alpha+1}$. However,
 since $J'$ does not qualitatively change our conclusions, we set $J'=0$ below \cite{SM}. 

\emph{Weak   coupling.---}Let us start with the weak-coupling limit, $|J_\chi|\ll J$. In the low-energy continuum limit and for decoupled chains, 
spin operators take the form ($x=ja$ with lattice constant $a$) \cite{Gogolin1998}
\begin{equation}\label{Sdef}
{\bf S}_\alpha(x) = {\bf J}_{L,\alpha}(x)+{\bf J}_{R,\alpha}(x)+(-1)^j {\bf n}_\alpha(x),
\end{equation}
where chiral spin currents ${\bf J}_{L/R,\alpha}(x)$ represent the smooth part
and ${\bf n}_\alpha(x)$ the staggered magnetization.  Using Abelian bosonization, we express these 
operators  in terms of   chiral bosons $\varphi_{L/R,\alpha}(x)$ or, equivalently,   dual fields  
$\phi_\alpha(x)=\left(\varphi_{L,\alpha}-\varphi_{R,\alpha}\right)/\sqrt2$ and
$\theta_\alpha(x)=\left(\varphi_{L,\alpha}+\varphi_{R,\alpha}\right)/\sqrt2$ \cite{Gogolin1998}.
With the non-universal constant $A\sim 1/a$   and $\nu=L/R=+/-$, one finds
\begin{eqnarray}
J_{\nu,\alpha}^z (x)&=& \frac{\nu}{\sqrt{4\pi}}\partial_x\varphi_{\nu,\alpha},
\quad J^\pm_{\nu,\alpha} (x)=\frac{1}{2\pi a}e^{\pm i\sqrt{4\pi} \varphi_{\nu,\alpha}}, \nonumber \\ 
\label{Sdef2}
n^z_\alpha(x) &=& A\sin[\sqrt{2\pi}\phi_\alpha ], \quad n_\alpha^\pm(x)=
A e^{\pm i\sqrt{2\pi}\theta_\alpha }.
\end{eqnarray}
For $J_\chi=0$,   open  boundary conditions at $x=0$ are imposed by 
writing 
$\varphi_{R,\alpha}(x)=\varphi_{L,\alpha}(-x)+ \varphi_0$  \cite{Gogolin1998}, where SU(2) invariance requires
$\varphi_0=0$ or $\varphi_0=\sqrt{\pi}$.
In terms of SU(2) currents,   we   have $\mathbf J_{R,\alpha}(x)=\mathbf J_{L,\alpha}(-x)$. 
The effective low-energy Hamiltonian can be written as $H_0 \simeq (2\pi v/3)\sum_\alpha\int_{-\infty}^{+\infty} dx\, \mathbf J_{L,\alpha}^2$, where $v\approx 1.17 Ja$ \cite{Eggert1992} is the spin   velocity for $J_2=0.2412J$. This model  has central charge $c=3$  corresponding to three decoupled SU(2)$_1$ Wess-Zumino-Novikov-Witten (WZNW) models \cite{Gogolin1998,Affleck1987}. 
We can then analyze the perturbations to the O point that arise for $|J_\chi|\ll J$. Boundary spin operators follow
 from Eq.~\eqref{Sdef} as ${\bf S}_\alpha \propto {\bf J}_{L,\alpha}(0)$ \cite{Eggert1992}. The three-spin   interaction  $\sim J_\chi {\bf J}_{L,1}(0)\cdot [{\bf J}_{L,2}(0)\times {\bf J}_{L,3}(0)]$ has scaling dimension three and is   irrelevant. In fact, it is more irrelevant than the leading $\mc T$-invariant perturbation $\sum_\alpha {\bf J}_{L,\alpha}(0)\cdot {\bf J}_{L,\alpha+1}(0)$ (dimension two), which is generated by the RG to second order in $J_\chi$. 
 
\emph{Strong   coupling.---}Next, we address the limit $|J_\chi| \gg J$.  For $J=0$,  one can readily  diagonalize the   three-spin Hamiltonian 
$H_{c}$   \cite{Wen1989}. The ground state of $H_c$ is twofold degenerate and, assuming $J_\chi>0$, has  eigenvalue $-\sqrt3/4$ of $\hat C$. In the $|S_1^z,S_2^z,S_3^z\rangle$ boundary spin basis, the ground state with  eigenvalue $M=+1/2$ of $\sum_\alpha S^z_\alpha$ is given by 
\begin{equation}\label{defplus}
|+\rangle = \frac{i}{\sqrt3}\left( |\downarrow\uparrow\uparrow\rangle
+ \omega |\uparrow\uparrow\downarrow\rangle+\omega^2 |\uparrow\downarrow\uparrow\rangle
\right),  \quad \omega=e^{2\pi i/3}.
\end{equation}
The $|-\rangle$ state with $M=-1/2$ follows by ${\cal P T}$ conjugation.
All other states involve an energy cost of order $J_\chi$. 
For finite $J\ll J_\chi$, the low-energy physics therefore involves an effective spin-$1/2$
 operator ${\bf S}_{\rm imp}$ acting in the $\{|+\rangle,|-\rangle\}$ subspace.
By projecting $H$ onto this subspace, we arrive at a spin-chain version of the three-channel Kondo model, 
\begin{equation}\label{3ck}
\tilde H  = H_0 + J_K {\bf S}_{\rm imp}\cdot\sum_\alpha
\left [{\bf   S}_{2,\alpha}+(J_2/J) {\bf   S}_{3,\alpha}\right],
\end{equation}
where $ J_K\simeq J/3$.  Since ${\bf S}_{\rm imp}$ is built from the original
boundary spins ${\bf S}_{j=1,\alpha}$, the latter disappear from $H_0$ and 
the boundary is now at site $j=2$. The exchange coupling $J_K$ is marginally relevant.
As a consequence, Kondo screening processes drive the system towards a strong-coupling
fixed point  identified with the   K point.   The physics of the K point is realized   at energy scales below the Kondo temperature $T_K\sim J e^{-1/\lambda_0}$, where  $\lambda_0 \approx J_K a/(2\pi v)$    \cite{Laflorencie2008}.  Although the projected Hamiltonian in Eq.~(\ref{3ck}) lacks $\mc T$-breaking interactions,  such interactions are   generated by a Schrieffer-Wolff transformation to first order in $J/J_\chi$. However, they turn out to
be irrelevant \cite{SM}. Before analyzing the K point using BCFT, we  turn to the   critical point separating the stable O and K points.
 
\emph{Chiral fixed point.---}We   define the   chirality
$\hat C_j=\mathbf S_{j,1}\cdot (\mathbf S_{j,2}\times \mathbf S_{j,3})$
  for three spins at  site $j$ in different chains, cf.~Eq.~(\ref{couplH}). 
  In the continuum limit, the most relevant contribution to $\hat C_j$ stems from the staggered magnetization, $\hat C_j\sim  \mathbf n_1(x)\cdot [\mathbf n_2(x)\times \mathbf n_3(x)]$. 
 Energetic considerations suggest that $J_\chi\neq0$ should  favor a fixed point in which $\hat C=\hat C_1$  acquires a   nonzero  expectation value. This happens if we impose  
\begin{equation}\label{cBC}
\varphi_{R,\alpha\pm 1}(x)=\varphi_{L,\alpha}(-x)+\varphi_0.
\end{equation}
As for the O point, SU(2) invariance requires  $\varphi_0=0$ or
$\varphi_0=\sqrt{\pi}$. Equation~\eqref{cBC}  
implements ideal chiral boundary conditions for the spin currents,
\begin{equation}\label{cBC2}
 \mathbf J_{R,\alpha\pm 1}(0)=\mathbf J_{L,\alpha}(0).
 \end{equation}  
 We refer to the corresponding fixed points as C$_\pm$, respectively. 
 
 \emph{Ideal spin circulator.---}To see that the C$_\pm$ points realize an ideal
 spin circulator, we 
 consider the linear spin conductance tensor (with arbitrary $y>0$ and $\omega\to i0^+$)  \cite{Meier2003,Oshikawa2006} 
\begin{equation}
\mathbb{G}_{\alpha\alpha'}^{bb'}  =   - \frac{(g\mu_{B})^{2}}{\hbar L\omega}
\intop_{0}^{L} dx \intop_{-\infty}^{\infty}d\tau\,e^{i\omega\tau } 
\left\langle {\cal T}_{\tau}J_{\alpha}^{b} (x,\tau )J_{\alpha'}^{b'} (y,0 ) \right\rangle,\label{eq:Kubo}
\end{equation}
which determines the spin current in chain $\alpha$ with polarization 
direction $\hat e_{b=x,y,z}$ in response to a spin chemical potential \cite{Jungwirth,Baltz2016} applied in chain $\alpha'$ with polarization $\hat e_{b'}$. 
Here $g$ denotes the gyromagnetic ratio, $\mu_B$ the Bohr magneton, $L$ the chain length, 
${\cal T}_\tau$ the imaginary-time ($\tau)$ ordering operator, and the spin current 
density is ${\bf J}_{\alpha}={\bf J}_{R,\alpha}-{\bf J}_{L,\alpha}$, cf.~Eq.~\eqref{Sdef}. Using the boundary conditions in  Eq.~(\ref{cBC2}), we obtain from Eq.~\eqref{eq:Kubo} the
maximally asymmetric tensor
\begin{equation}\label{ssc}
\mathbb{G}_{\alpha\alpha'}^{bb'} =\frac{(g\mu_B)^2}{2\pi\hbar}\delta^{bb'}
\left (\delta_{\alpha,\alpha'}-\delta_{\alpha\pm1,\alpha'} \right) \qquad \text{(for C$_\pm$)}.
\end{equation}
Right at the C$_+$ or C$_-$ point, an incoming spin current is therefore completely channeled into the adjacent chain $\alpha\pm 1$, cf.~Fig.~\ref{fig1}, without polarization change.  The Y junction then represents an ideal spin circulator.

\textit{Realizing the chiral point.---}It remains to  show that  the C$_\pm$ points can be realized at intermediate $J_\chi$. We first approach the problem from the weak coupling side. Despite being energetically favored by $J_\chi\neq 0$, the C$_\pm$ points must be unstable since the O point is stable for $|J_\chi|\ll J$.  Indeed, a relevant boundary perturbation, $H_1$, is generated by the three-spin coupling when using Eq.~(\ref{Sdef}) and imposing either of the conditions (\ref{cBC}),
\be \label{H1}
H_1=\lambda_1 \sum_\alpha \cos\left\{ \sqrt{\pi} \left[\varphi_{L,\alpha}(0)-
\varphi_{L,\alpha+1}(0)\right] \right\}.
\ee
Using bosonization, we find $\lambda_1<0$ and $|\lambda_1|\propto |J_\chi|$ for $|J_\chi|\ll J$.
The  physical process  behind t
his  dimension-$1/2$  operator is  the backscattering of spin currents  \cite{Oshikawa2006}. 
For $\lambda_1<0$, the RG flow approaches $\lambda_1\to-\infty$ at low energies. Pinning the 
boson fields to the respective cosine minima in Eq.~\eqref{H1} takes the system back to the   O point. 
Since at weak coupling there is only one relevant perturbation allowed by $\mathbb{Z}_3$  symmetry, the C point can be reached by fine tuning a single parameter $\lambda_1$, e.g., by increasing $J_\chi$.
Let us assume that there is a critical value $J_\chi^c$ such that $\lambda_1(J_\chi^c)=0$. For $J_\chi>0$ ($J_\chi<0$), this putative critical point corresponds to the C$_-$ (C$_+$) point. 

Now consider approaching the C point from the strong coupling side. For $J_\chi>J_\chi^c$, the relevant coupling constant becomes positive, $\lambda_1>0$,  and the RG flow approaches $\lambda_1\to +\infty$.  The pinning conditions 
now involve a $\pi$-phase shift for the cosine terms in Eq.~\eqref{H1} as compared
to  $J_\chi<J_\chi^c$.  For the total magnetization 
$S^z_{\rm tot}=-\sum_\alpha[\varphi_{L,\alpha}(0)-\varphi_{R,\alpha}(0)]/\sqrt{2\pi}$, this shift means that an effective spin-$1/2$ degree of freedom 
has been brought from infinity to the boundary. This is precisely what we expect from the formation of the impurity spin   in the strong coupling regime. 
The coupling of the   impurity spin to the bulk allows for a second dimension-$1/2$ boundary operator,   $H_2=\lambda_2\mathbf S_{\text{imp}}\cdot \sum_\alpha \tilde{\mathbf n}_\alpha(0)$, where $\tilde{\mathbf n}_\alpha$ is the staggered magnetization after imposing Eq.~(\ref{cBC}).
The flow of $\lambda_1$ and $\lambda_2$ to strong coupling leads to a fixed point where the impurity spin is overscreened by the three chains, which we identify with the K point. 
 
Since $\lambda_1$  vanishes at the critical point, the effects of the dimension-$1/2$
perturbations  are felt only when the renormalized couplings at energy scale ${\cal E}$ become of order one. We thus obtain a wide quantum critical regime, ${(1-J_\chi/J_\chi^c)^2}\lesssim{\cal E}/J\ll 1$, where the physics is governed by the C point. Related but different
chiral points have been discussed for electronic Y junctions \cite{Oshikawa2006}. The latter are 
stable for attractive electron-electron interactions and the asymmetry of the charge conductance tensor depends on the interaction strength. By contrast, our C point is unstable, but due to SU(2) symmetry the spin conductance \eqref{ssc} is universal and maximally asymmetric. 

\emph{BCFT approach.---}A spin-$1/2$ impurity coupled with equal strength to the open ends of two spin chains realizes a spin version of the two-channel Kondo effect    \cite{Eggert1992,Affleck1993,Alkurtass2016}. Here we develop a BCFT approach and extend this analogy to three channels. We employ the conformal embedding 
SU(2)$_3\times \mathbb Z_3^{(5)}$, whereby the total central charge $c=3$ is split into a SU(2)$_3$ WZNW model (with $c=9/5$), representing the  spin degree of freedom, and 
a parafermionic $\mathbb Z_3^{(5)}$ CFT (with $c=6/5$) \cite{Zamolodchikov1985,Fateev1987,Frenkel1992,Totsuka1996,Affleck2001},   representing the   ``flavor'' (i.e., channel) degree of freedom. 

The RG  fixed  points are characterized by conformally invariant boundary conditions 
 \cite{Cardy1986,Cardy1989,Affleck1991}. The spectrum of the theory is encoded by the partition function $Z_{\text{AB}}$ on the cylinder with boundary conditions A and B.  For instance, $Z_{\text{OO}}$ represents the partition function with open boundary conditions at both ends.  Partition functions with other boundary conditions can be generated via fusion \cite{Cardy1989}. The boundary operators that perturb the K point can be determined using  double fusion with the spin-$1/2$ primary in the SU(2)$_3$ sector  \cite{AffleckLudwig2,Affleck1993}. 
 The leading irrelevant operator is the Kac-Moody descendant $\boldsymbol{\mc J}_{-1}\cdot \boldsymbol{\phi}_1$, where $\boldsymbol{\mc J}$ is the SU$(2)_3$ current and $\boldsymbol{\phi}_1$ is the spin-$1$ primary. This $\mc T$-invariant operator has scaling dimension $\Delta=7/5$,  as in  the free-electron three-channel Kondo model \cite{Affleck1991,AffleckLudwig2}.
Similarly,  the leading chiral boundary operator at the K
point is the   dimension-$8/5$ field of $\mathbb Z_3^{(5)}$ \cite{Fateev1987}.
Moreover, the effective Hamiltonian   at the K point includes only irrelevant boundary operators
in the presence of cyclic exchange symmetry \cite{SM}.

\begin{figure}[t]
\centering
\includegraphics[width=0.83\columnwidth]{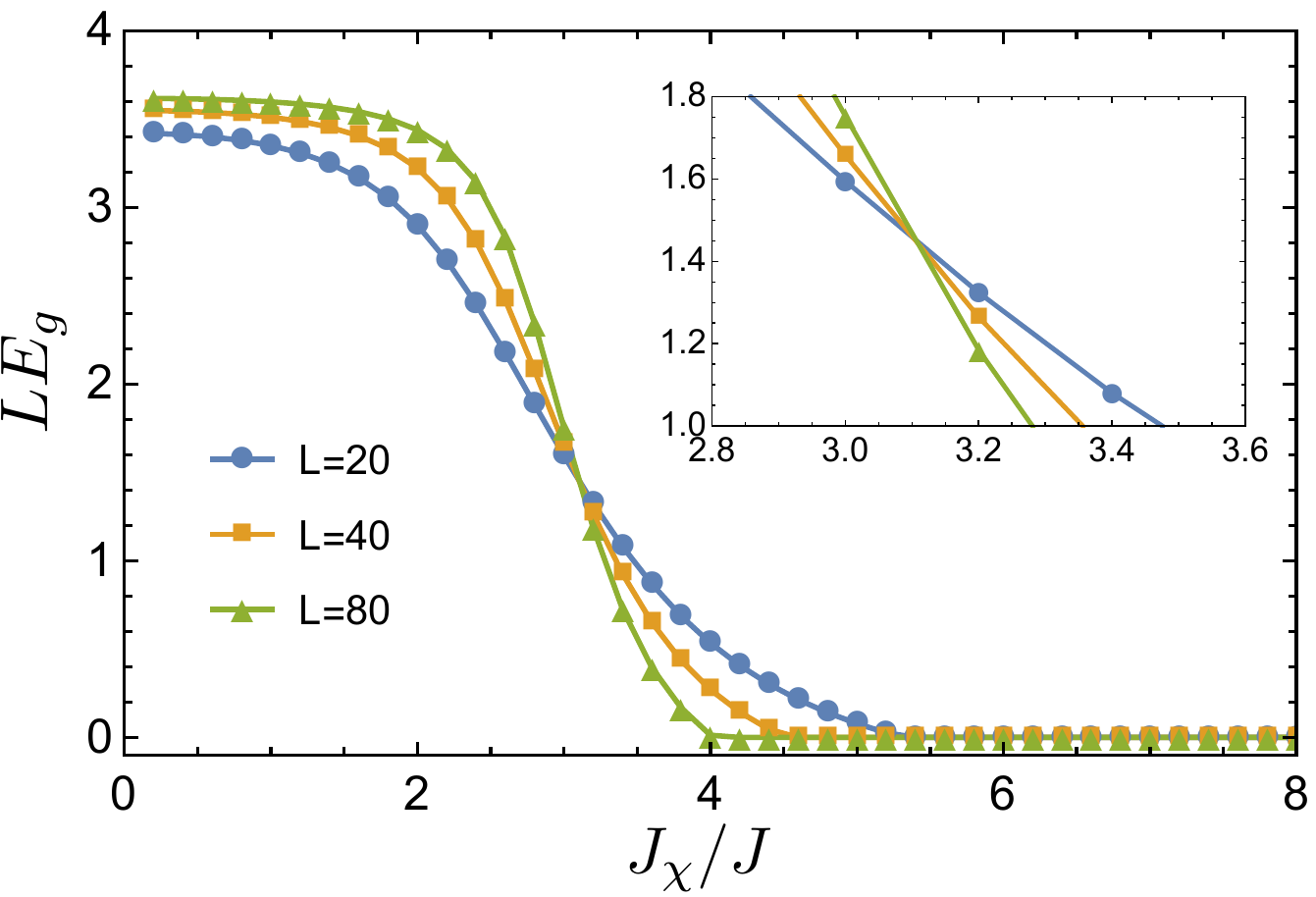}
\caption{DMRG results for the finite-size energy gap $E_g$, rescaled by the chain length $L$, vs $J_\chi/J$ for several $L$. The inset highlights the crossing that determines the critical point.  }
\label{fig2}
\end{figure} 

\begin{figure}[t]
\centering
\includegraphics[width=0.9\columnwidth]{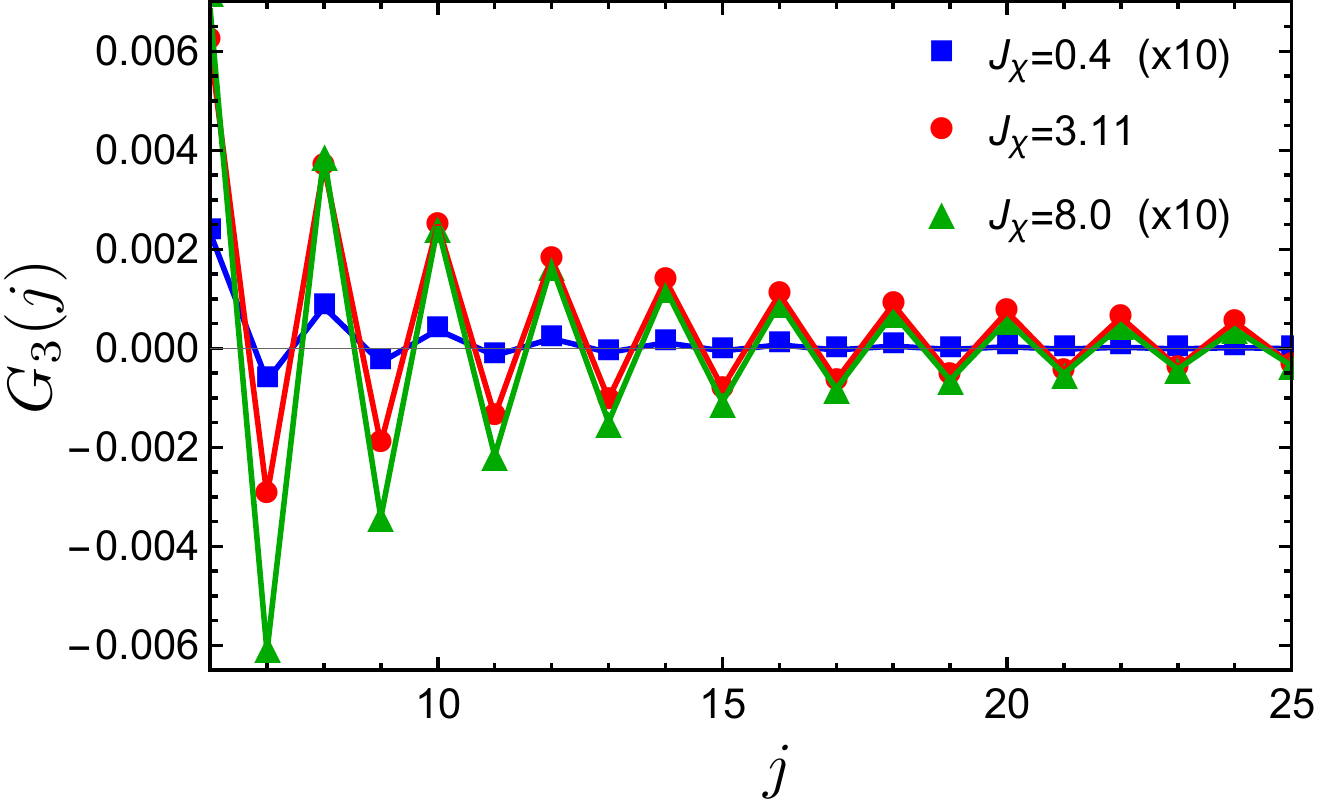}
\caption{Three-spin correlations, $G_3(j)$, vs distance from the junction for $L=80$ and three values of $J_\chi$. The data for $J_\chi=0.4\,J$ and $J_\chi=8\,J$ are scaled up by a factor 10.  Solid lines represent fits to a power-law decay.}
\label{fig3}
\end{figure} 

 \begin{table}[t]%The best place to locate the table environment is directly after its first reference in text
\caption{%
Exponent  $\nu(J_\chi)$ obtained by fitting  the decay of $G_3(j)$ in the interval $8\leq j\leq L/2$. The extrapolated value follows from a second-order-polynomial fit. The last column shows the predictions for the O, C, and K points, respectively. \label{exponents}
}
\begin{tabular}{ c @{\quad} c @{\quad} c @{\quad}c@{\quad} c@{\quad}c@{\quad} c}
\toprule
$J_\chi/J$&&
$L=40$& $L=60$&$L=80$&Extrap.&Expected\\
\colrule
$0.4$ & &$3.56$&$3.51$&$3.49$&$3.45$&$3.5$\\
$3.11$ & &$1.89$&$1.79$&$1.74$&$1.59$&$1.5$\\
$8$ & & $2.31$&$2.22$&$2.18$&$2.08$&$2.1$\\
\botrule
\end{tabular}
\end{table} 

\emph{DMRG results.---}We now  describe numerical results for Y junctions with chain length $L$ using the DMRG algorithm by Guo and White \cite{Guo2006}, which works efficiently for open boundary conditions at  $j=L$. First, we look for the critical point by analyzing the finite-size gap $E_g$ between the lowest-energy state with  $S^z_{\text{tot}}=\sum_{j,\alpha}S^z_{j,\alpha}=0$ and the one with $S^z_{\text{tot}}=1$. For large $L$, at weak coupling we expect $E_g$
to approach the  singlet-triplet  gap of decoupled chains (O point),
 $E_g = \pi v/L$. On the other hand, at strong coupling, the BCFT approach predicts (through the partition function   $Z_{\text{KO}}$   \cite{SM}) that the ground state is a triplet and hence $E_g$ should vanish 
 identically. We indeed observe a ($L$-dependent) level crossing between a singlet ground state for small $J_\chi$ and a triplet  for large $J_\chi$, see Fig.~\ref{fig2}. The critical point is then determined from the crossing of the $L E_g$ vs $J_\chi$ curves for $40\leq L\leq80$, resulting in 
 $J_\chi^c/J=3.11(1)$.

Next, we calculate the three-spin ground-state correlation function
$G_3(j)=\langle  \hat C_j\rangle=\langle\mathbf S_{j,1}\cdot (\mathbf S_{j,2}\times \mathbf S_{j,3})\rangle$.  At the C point, the long-distance decay of $G_3(j)$ is governed by the bulk scaling dimension of $\hat C_j$, where our BCFT predicts $G_3(j)\sim (-1)^jj^{-\nu_\text{C}}$ with  $\nu_\text{C}=3/2$. Near the $\mc T$-symmetric O and K points, the leading chiral boundary operator has dimension $\Delta_{\rm O}=3$ and $\Delta_{\rm K} =8/5$, respectively. 
Standard perturbation theory around these fixed points     yields $G_3(j)\sim (-1)^j j^{-\nu_{\text{O,K}}}$ with $\nu_\text{O}=7/2$ and $\nu_\text{K}=21/10$, respectively.
Our DMRG results for $G_3(j)$ are shown in Fig.~\ref{fig3}. First, we note that $G_3(j)$ has much larger magnitude and decays more slowly at the critical point. Fitting the numerical results to a power law expression with smooth and  staggered parts yields the exponent $\nu(J_\chi)$ of the dominant staggered term as listed in Table \ref{exponents}.  For the fit, we only took into account  data for $G_3(j)$ with $8\leq j\leq L/2$ in order to avoid both the non-universal short-distance behavior and  effects due to the open boundary at $j=L$.  (Results for $\nu(J_\chi)$ are robust under 
changes of the fitting interval \cite{SM}.) 
Our DMRG results in Table \ref{exponents} agree well with the analytical predictions. The deviation is most significant   at the C point, where one however also observes the strongest finite-size effects.   We  emphasize that the DMRG results show a slow decay of $G_3(j)$  over a wide region around the critical point. 
 
\emph{Conclusions and Outlook.---}We have demonstrated that a Y junction of Heisenberg chains acts as a   quantum spin circulator in the vicinity of a critical point  reached  by tuning the  three-spin interaction $J_\chi$. In addition to applications as a nonreciprocal device for pure spin transport, this spin circulator can be used for constructing two-dimensional networks realizing CSL phases, where the chirality of each node can be independently tuned \cite{Claassen2017}.  In fact, such an approach could allow for the systematic design of  synthetic quantum materials harboring CSL phases. For 
instance, the network with uniform chirality shown in Fig.~\ref{fig1}(b) has  spin modes circulating in closed loops in the bulk. The bulk quasiparticles can be defined from the spin-1/2 field  of the  chiral WZNW model in each loop \cite{Gorohovsky2015} and have a finite gap   due to the finite length of the loops. In addition, there is a gapless chiral edge mode with quantized spin conductance, cf.~Fig.~\ref{fig1}(b). This corresponds to the properties of the Kalmeyer-Laughlin CSL, a topological phase equivalent to a bosonic fractional quantum Hall system \cite{Kalmeyer1987,Bauer2014}. Furthermore, one can consider  networks with alternating sign of $J_\chi$, i.e., staggered chirality between the nodes. This may  shed light on the much less understood gapless CSLs with spinon Fermi surfaces \cite{Bauer2013arXiv,Bieri2015}. 

\begin{acknowledgments} 
We thank I. Affleck, E. Ercolessi, F. Ravanini, A. Tsvelik, and J.C. Xavier for discussions. We acknowledge funding by the Deutsche Forschungsgemeinschaft within the network CRC TR 183 (project C01) and by CNPq (R.G.P.).
\end{acknowledgments}

\bibliographystyle{apsrev4-1}

\bibliography{chiral}

\onecolumngrid

\appendix
%\maketitle

\section{Supplemental Material for ``Quantum  spin circulator in  Y junctions of Heisenberg chains'' }

\subsection{1. Effective Hamiltonian in the strong coupling limit}

We consider the Hamiltonian for three boundary spins:\be
H_c=J'\sum_{\alpha}\mathbf S_{\alpha}\cdot \mathbf S_{\alpha+1}+J_\chi \mathbf S_{1} \cdot (\mathbf S_{2} \times \mathbf S_{3} ).
\ee
Here we have included the exchange coupling $J'$ for a more general discussion. Eigenstates of $H_c$ are  labeled by:  (i) the total boundary spin $S=1/2$ or $S=3/2$, (ii)  the magnetic quantum number  $M=-S,\dots,S$, and (iii) the eigenvalue  $\sigma \sqrt 3 /4$ of the scalar spin chirality $\hat C$. The spin chirality vanishes in the fourfold degenerate $S=3/2$ sector ($\sigma=0$). The $S=1/2$  sector splits into two doublets with opposite chirality $\sigma=\pm1$. The   energies are 
 \bea
E\left(S=\frac32,\sigma=0\right)&=&\frac{3J^\prime}4,\\
E\left(S=\frac12,\sigma=\pm1\right)&=&-\frac{3J^\prime}4\pm \frac{\sqrt3}{4}J_\chi.
\eea
The ground state is   twofold degenerate for $J'\geq 0$. For $J_\chi>0$, the two ground states   are the negative-chirality states \bea
|+\rangle&=&\left|S=\frac12,M=+\frac12,\sigma=-\right\rangle=\frac{i}{\sqrt3}[|\downarrow\uparrow\uparrow\rangle+\omega|\uparrow\uparrow\downarrow\rangle+\omega^2|\uparrow\downarrow\uparrow\rangle],\\
|-\rangle&=&\left|S=\frac12,M=-\frac12,\sigma=-\right\rangle =-\frac{i}{\sqrt3}[|\uparrow\downarrow\downarrow\rangle+\omega^2|\downarrow\uparrow\downarrow\rangle+\omega|\downarrow\downarrow\uparrow\rangle],\qquad \omega=e^{i2\pi/3}.\label{chiralstates}
\eea  

We can   treat the coupling  to the chains  in the   strong coupling limit $|J'|,|J_\chi|\ll J$   using  degenerate perturbation theory.  Let us define $\hat P_\sigma$ as the projector onto the subspace of states with chirality $\sigma$. 
%The projection of the impurity spins onto the low-energy doublet yields \be
%\hat P_-\mathbf S_\alpha  \hat P_-=\frac{1}3\mathbf S_{\text{imp}},
%\ee
%where  $\mathbf S_{\text{imp}}=(\sigma^x,\sigma^y,\sigma^z)/2$ is the effective impurity spin with Pauli matrices acting in the low-energy subspace   $\{|+\rangle,|-\rangle\}$. The first term in the effective Hamiltonian is simply the projection  \bea
%H_{\text{eff}}^{(1)}&=&\hat P_-H_0\hat P_-\nonumber\\
%&=&\sum_\alpha\sum_{j\geq2}(J\mathbf S_{j,\alpha}\cdot\mathbf S_{j+1,\alpha}+J_2\mathbf S_{j,\alpha}\cdot\mathbf S_{j+2,\alpha})+\frac{1}{3}\mathbf S_\text{imp}  \cdot \sum_\alpha (J\mathbf S_{2,\alpha}+J_2\mathbf S_{3,\alpha}).
%\eea
We then compute the effective Hamiltonian up to second order in $J/J_\chi$.
% using  \be
%H_{\text{eff}}^{(2)}=\hat P_-\hat V (E_0-H_c)^{-1}(\hat P_0+\hat P_+)\hat V\hat P_-,
%\ee
%where $\hat V= \sum_{\alpha}\mathbf S_{1,\alpha}\cdot (J \mathbf S_{2,\alpha}+J_2 \mathbf S_{3,\alpha})$ is the perturbation and $E_0=- ({3J^\prime}+{\sqrt3}J_\chi)/4$ is the unperturbed ground state energy.
  The result is\bea
H_{\text{eff}}&=&\sum_\alpha\sum_{j\geq2}(J\mathbf S_{j,\alpha}\cdot\mathbf S_{j+1,\alpha}+J_2\mathbf S_{j,\alpha}\cdot\mathbf S_{j+2,\alpha})+J_K\mathbf S_{\text{imp}}  \cdot \sum_\alpha \tilde {\mathbf S}_{b,\alpha}\nonumber\\
&&+\tilde{J}^{\prime}\sum_{\alpha}\tilde {\mathbf S}_{b,\alpha}\cdot\tilde {\mathbf S}_{b,\alpha+1}+\tilde{J}_{\chi}\mathbf S_{\text{imp}}\cdot \sum_\alpha\tilde {\mathbf S}_{b,\alpha}\times\tilde {\mathbf S}_{b,\alpha+1},\label{effectiveH}
\eea
where  $\tilde {\mathbf S}_{b,\alpha}=\mathbf S_{2,\alpha }+(J_2/J)\mathbf S_{3,\alpha}$ and
%\bea
%J_K&=&\frac{J}3-\frac{J^2}{72J_\chi}\frac{5\sqrt3-2r}{1+2r/\sqrt3},\\
%\tilde{J}^\prime&=&\frac{J^2}{72J_\chi}\frac{3\sqrt3-2r}{1+2r/\sqrt3},\\
%\tilde{J}_\chi&=&\frac{J^2}{12J_\chi}\frac{1}{1+2r/\sqrt3},
%\eea
\be
J_K=\frac{J}3-\frac{J^2}{72J_\chi}\frac{5\sqrt3-2r}{1+2r/\sqrt3},
\qquad \tilde{J}^\prime=\frac{J^2}{72J_\chi}\frac{3\sqrt3-2r}{1+2r/\sqrt3},
\qquad \tilde{J}_\chi=\frac{J^2}{12J_\chi}\frac{1}{1+2r/\sqrt3}
\ee
with $r=J'/J_\chi$.  Note that in this limit the  $\mc T$-breaking perturbation  $\tilde{J}_\chi$ appears at order $J^2/J_\chi$. At energy scales $T_K<{\cal E}\ll J_\chi$, we can treat  the boundary couplings  perturbatively and take the continuum limit in the form $\mathbf S_{b,\alpha} \propto \mathbf J_{L,\alpha}(0)$. In this case, the couplings generated by $\tilde J'$ and $\tilde J_\chi$ are   irrelevant. The Kondo coupling $J_K>0$ is marginally relevant and drives the system to the K fixed point at energy scales ${\cal E}\ll T_K$. Remarkably, the effective Hamiltonian in the strong coupling limit is valid for arbitrary $r\geq 0$, implying that the existence of the K fixed point at strong coupling (and of a critical point separating it from the O point at weak coupling) is not particular  to   $J'=0$.

\subsection{2. Boundary operators in the  boundary conformal field theory approach }

The model of three decoupled spin chains has  total central charge $c=3$ and  a global SU$(2)\times$SU$(2)\times$SU$(2)$ symmetry. The currents ${\bf J}_{L,\alpha}$ generating the SU$(2)_1$ symmetry for each spin chain ($\alpha=1,2,3$) have dimension $1$ and are characterized by the operator product expansion (OPE) \cite{di1997conformal} 
\begin{equation}
J^{a}_{L,\alpha}\left(z\right) J^{b}_{L,\alpha} \left(w\right) \sim\frac{\delta^{ab}}{\left(z-w\right)^{2}} +\frac{i\epsilon^{abc}}{z-w}J^{c}_{L,\alpha}\left(w\right),\label{KMope}
\end{equation}
while they simply commute for different legs. Here, we use $z=v\tau+ix$ and $\bar z=v\tau-ix$.  %satisfies the $SU(2)_{k}$ Kac-Moody (KM) algebra
Useful linear combinations of these currents are the ``helical''
currents% \cite{Ingersent2005} 
\begin{eqnarray}
&  & \boldsymbol{\mc I}_{h} =\sum_{\alpha=1}^{3}\omega^{\alpha\,h}\mathbf{J}_{\alpha},\qquad\qquad\omega=e^{\frac{2\pi i}{3}}\label{helicalcurrent},
\end{eqnarray}
with $h=-1,0,1$. The latter satisfy the OPE
\be
\mc{I}_{h}^a\left(z\right)\mc{I}_{h'}^b
\left(w\right)\sim\frac{3\delta_{h+h'}\delta^{ab }}{\left(z-w\right)^{2}}+\frac{i
\epsilon^{abc}}{z-w} \mc{I}_{h+h'}^c \left(w\right)\label{HelicalCurrentOPE},
\ee
where the sum $h+h'$ is defined modulo $3$. Note that $\boldsymbol{\mc I}_{0}$ is the level-$3$ current $\boldsymbol{\mc J}$ of the main text.

The only (marginally) relevant interaction in the strong-coupling Hamiltonian \eqref{effectiveH} is the Kondo term, with coefficient $J_K$, which couples the impurity spin to the level-$3$ current $\boldsymbol{\mc J}$. This selects the SU$(2)_3$ WZW conformal field theory with central charge $9/5$ as part of our embedding.
This theory possesses a finite number of primary fields $\boldsymbol{\phi}_s$, corresponding to integrable representations of SU$(2)$, labeled by the spin $s=0,1/2,1,3/2$. The corresponding scaling dimensions are $\Delta_s=s(s+1)/5$ and the primary fields obey the fusion rules
\be \label{phiphi}
\boldsymbol{\phi}_s\times \boldsymbol{\phi}_{s'} = \boldsymbol{\phi}_{\left|s-s'\right| } + \boldsymbol{\phi}_{\left|s-s'\right|+1 } + \ldots + \boldsymbol{\phi}_{\min\left\{s+s',3/2\right\} },
\ee
while their OPE with the currents is
\be \label{Jphi}
\boldsymbol{\mc J}^{a} \left(z\right) 
\boldsymbol{\phi}_s\left(w\right) \sim -
\frac{1}{z-w} T^{a}_s \boldsymbol{\phi}_s\left(w\right),
\ee
with
$T^{a}_s$ being the element $a$ of the SU$(2)$ generators in the spin-$s$
representation  \cite{di1997conformal}.
% \be 
%\Delta_0=0 , \quad \Delta_{1/2}=\frac3{20},\quad \Delta_1=\frac25, %\quad \Delta_{3/2}=\frac34.\ee

The remaining central charge $3-9/5=6/5$ must be associated with the ``flavor''   degree of freedom. Hence, possible  conformal embeddings are: (i) SU$(2)_3\times \mc M_{3,4}\times \mc M_{4,5}$, where the minimal model $ \mc M_{3,4}$ is the Ising model and  $\mc M_{4,5}$ is the tricritical Ising model; (ii) SU$(2)_3\times \mathbb Z_3^{(5)}$, where $\mathbb Z_{3}^{(5)}$ is a field theory with % an intrinsic  $\mathbb Z_3$ symmetry and
 an infinite-dimensional symmetry (called $W$ symmetry) in addition to conformal symmetry \cite{Zamolodchikov1985,Fateev1987,Totsuka1996}. 
%The relation between the $\mathbb Z_3^{(5)}$ theory and the product of Ising and tricritical Ising models has been  discussed in \cite{Affleck2001}.
 The two embeddings can generate nonequivalent sets of possible boundary conditions via fusion \cite{Affleck2001}. Only the embedding (ii) allows us to reproduce the boundary conditions in Eq. (6) of the main text, derived using abelian bosonization.  Primary fields then have scaling dimension $\Delta=\Delta_s+\Delta_f$, where $\Delta_s$ $\left(\Delta_f\right)$ is the dimension of the primary field in the SU(2)$_3$  $\left(\mathbb Z_3^{(5)}\right)$ sector.

The $\mathbb Z_3^{(5)}$ theory has 20 primary fields. Most important for our purposes are the operators $\Psi,\Psi^*,\Omega,\zeta,\zeta^*,\varepsilon,\varepsilon'$, having conformal dimensions $
\Delta_{\Psi}=\Delta_{\Psi^*}=\frac35,%\quad
\Delta_\Omega=\frac85,%\quad
\Delta_{\zeta}=\Delta_{\zeta^*}=2,%\quad
\Delta_{\varepsilon}=\frac1{10},%\quad
\Delta_{\varepsilon'}=\frac1{2}.\label{dimZ3}
$
The operator algebra has been computed in \cite{Frenkel1992,Totsuka1996}. We note in particular the fusion rules:
\be
\Psi\times \Psi^*=\mathbb I+\Omega,\qquad \Psi\times \Psi=\Psi^*,\qquad\Psi\times\Omega=\Psi+\zeta^*.\label{fusionZ3}
\ee

 \begin{table}[t]%The best place to locate the table environment is directly after its first reference in text
\caption{%
Operator content in the cylinder partition function  $Z_{\text{AB}}$ for  boundary conditions A and B. We  consider chains with an even number of sites and integer total spin. We  separate singlet   ($s=0$) from triplet ($s=1$) states. The numbers  represent the dimension $\Delta_f$ of the primary fields in the $\mathbb Z_3^{(5)}$ CFT; 
%repeated entries indicate that there are two independent primary fields with the same dimension.
in brackets, the number of primary fields with the same dimension.\label{dimensions}
}
\begin{tabular}{ l @{\qquad} c c c c}
\toprule
\textrm{AB}&&
$s=0 \,(\Delta_s=0)$& &$s=1 \,(\Delta_s=2/5)$\\
\colrule
OO & &$0,2(\times2)$& &$3/5(\times2),8/5$\\
KK & &$0,3/5(\times2),8/5,2(\times2)$& &$0,3/5(\times2),8/5,2(\times2)$\\
CC & & $0,1/2(\times3),2(\times2)$& &$1/10(\times3),3/5(\times2),8/5$\\
KO & & $3/5(\times2),8/5$& &$0,3/5(\times2),8/5,2(\times2)$\\
\botrule
\end{tabular}
\end{table} 

We now identify the original SU(2)$_1$ currents $\mathbf J_{\alpha}$   in terms of the operator content of SU$(2)_3\times \mathbb Z_3^{(5)}$. Besides the SU$(2)_3$ current $\boldsymbol{\mc I}_{0}$, which we have already written as the sum of SU$(2)_1$ currents, the only dimension-$1$ operators that we can construct are $\boldsymbol \phi_1\Psi$ and $\boldsymbol \phi_1\Psi^*$. Comparing Eqs.~\eqref{phiphi}, \eqref{Jphi} and \eqref{fusionZ3} with Eq.~\eqref{HelicalCurrentOPE}, we conclude that
\be
\boldsymbol \phi_1\Psi \sim \boldsymbol{\mc I}_{1},
\qquad\qquad
\boldsymbol \phi_1\Psi^* \sim \boldsymbol{\mc I}_{-1}.
\ee
From the above relation, we infer that the cyclic exchange $\alpha\to\alpha-1$ acts nontrivially   in the $\mathbb Z_3^{(5)}$ sector as $\Psi\to \omega \Psi$, $\Psi^*\to \omega^* \Psi^*$.
%\be
%\Psi\to \omega \Psi,\qquad \Psi^*\to \omega^* \Psi^*.\label{Z3Psi}
%\ee
Therefore, $\Psi$ and $\Psi^*$ are not invariant under cyclic exchange.  Moreover, $\mc P$ exchanges $\Psi$ and $\Psi^*$. Time reversal acts nontrivially in the SU(2)$_3$ sector, flipping the sign of spinful fields. In addition, $\mc T$ involves complex conjugation, in particular exchange of right and left movers, $z\leftrightarrow \bar z$.

Let us now consider the scalar spin chirality operator $\hat C_j=\epsilon^{abc}S^a_{j,1}S^b_{j,2}S^c_{j,3}$. Substituting the expansion for the spin operators in Eq. (2) of the main text, we find that  the slowest decaying component stems from the staggered magnetization in all three chains: $\hat C_j\sim (-1)^j\epsilon^{abc}n^a_1(x)n^b_2(x)n^c_3(x)$. This scalar operator has scaling dimension $3/2$ and zero conformal spin. The counterpart in the embedding SU$(2)_3\times \mathbb Z_3^{(5)}$ must be a linear combination of the operators with the same scaling dimension. The actual combination can be fixed by imposing that the operator be invariant under cyclic exchange and odd under $\mc P$ and $\mc T$. We must then have
\be
\hat C_{j}(\tau)\sim i\text{tr}[\boldsymbol{\phi}_{1/2}(z)\otimes \boldsymbol{\phi}_{1/2}(\bar z)][\Psi(z)\Psi^*(\bar z)-\Psi(z)\Psi^*(\bar z)], 
\label{Chiralopembed}
\ee
for $x=ja$  in the bulk (i.e.  far from the boundary). To obtain the chirality at the boundary, $\hat C=\hat C_1$, we take the boundary limit $x\to 0$ in  Eq. (\ref{Chiralopembed}) using the fusion rules  for the SU(2)$_3$ WZNW model and $\mathbb Z_3^{(5)}$ theory. The leading operator generated by the OPEs is\be
\hat C(\tau)\sim \Omega(\tau),
\ee
i.e., the chirality at the boundary is represented by the dimension-$8/5$ primary field of $\mathbb Z_3^{(5)}$. In fact, this  operator appears in the partition function on the cylinder with Kondo boundary conditions at both ends, $Z_{\text{KK}}$, see Table \ref{dimensions}.  The latter is  obtained from the partition function with open boundary conditions, $Z_{\text{OO}}$, by double fusion with the spin-$1/2$ primary   in the  SU(2)$_3$ sector. Therefore, the dimension-$8/5$ boundary operator  is an allowed perturbation to the three-channel Kondo fixed point if $\mc T$-symmetry is broken but $\mc P\mc T$ and $\mathbb Z_3$ symmetries are preserved.  We note that the relevant (dimension-$3/5$) operators $\Psi$ and $\Psi^*$ also  appear in $Z_{\text{KK}}$, but they are not allowed in the Hamiltonian as long as the $\mathbb Z_3$ cyclic exchange symmetry is preserved. %At the O point, the leading chiral boundary operator is  simply the triple product of   decoupled SU(2)$_1$ currents, $\mathbf J_{L,1}(0)\cdot [\mathbf J_{L,2}(0)\times \mathbf J_{L,3}(0)]$. 

We have also obtained the C$_\pm$ fixed points by fusion with either of the two dimension-$1/9$ primary fields in the $\mathbb Z_3^{(5)}$ sector \cite{Fateev1987}. The boundary operators that perturb the C$_\pm$ points can be read off from  $Z_{\text{CC}}$  in Table \ref{dimensions}. The sum of the $s=0$ dimension-$1/2$ operators is identified with the $\lambda_1$ perturbation  in Eq. (9) of the main text. For $J_\chi>J_\chi^c$, the   $s=1$ operator with $\Delta=\frac25+\frac1{10}$ can be combined with $\mathbf S_{\text{imp}}$ to produce the $\lambda_2$ perturbation.

\subsection{3. DMRG methods}

In order to study the Y junction with the chiral boundary interaction in this work,  we have used a suitable extension
of the DMRG proposed by Guo and White \cite{Guo2006}. It is possible to use the ordinary
DMRG \cite{White1992} to investigate such junctions by mapping the junction to a one-dimensional
system with long-range interactions. However,
the computational effort required to treat these interactions is equivalent
to considering periodic boundary conditions. Therefore,  a large
truncated Hilbert space is necessary in order to obtain results with a reasonable
accuracy. In contrast, the accuracy achieved
by the  procedure of Ref. \cite{Guo2006}  is close to that of 
DMRG  for open chains.

Using the DMRG to estimate the energies and the three-spin correlations
of finite-size Y junctions, we have considered up to $m=150$ kept
states per block. At the final sweep, the truncation error
is typically smaller than $10^{-8}$.
In order to check the accuracy of our DMRG results, for fixed system size, we compared the numerical data obtained by keeping $m\approx150$ and $m=50$ states. We have observed that the energies are obtained with a precision of at least $\sim10^{-5}$.  Finally, we emphasize that errors related to
the choice of fitting interval in determining the power-law exponent $\nu(J_x)$ of three-spin correlations, cf.~Table I in the main text, are at least one order of magnitude smaller than the 
values acquired by the DMRG.  
In Table \ref{tableSM}, we summarize our DMRG results  for different fitting intervals 
in order to validate this statement.

 \begin{table}[t]%The best place to locate the table environment is directly after its first reference in text
\caption{%
Exponents $\nu(J_\chi)$ for the power-law decay of $G_3(j)$ as listed
in Table I of the main text but using different fitting intervals.\label{tableSM}
}
\begin{tabular}{c @{\quad} c @{\quad} c @{\quad} c @{\quad}c@{\quad} c}
\toprule
                 Interval&        $J_\chi/J$&         $L=40$&   $L=60$&  $L=80$\\ 
\colrule
                               &                 $0.4$&         $3.56$&       $3.51$&       $3.49$\\
$8\leq j\leq L/2$ &               $3.11$&         $1.89$&       $1.79$&        $1.74$\\
                               &                    $8$&         $2.31$&       $2.22$&        $2.18$\\
                               &                 $0.4$&                  --&    $3.50$&        $3.49$\\
$6\leq j\leq L/3$ &                $3.11$&                 --&   $1.77$&          $1.74$\\
                               &                      $8$&                 --&   $2.20$&          $2.18$\\
                                &                  $0.4$&                --&   $3.52$&            $3.49$\\
$8\leq j\leq L-20$&                $3.11$&                --&   $1.83$&            $1.78$\\
                                &                      $8$&               --&   $2.24$&             $2.2$\\
\botrule
\end{tabular}
\end{table}

\end{document}